\documentclass[prd,aps,twocolumn,showpacs,preprintnumbers,amsmath,amssymb]{revtex4} 
\usepackage{graphicx}% Include figure files 
\usepackage[]{caption} 
\usepackage{amsmath} 
\usepackage{amssymb} 
 
\def\be{\begin{equation}} 
\def\ee{\end{equation}} 
\def\bea{\begin{eqnarray}} 
\def\eea{\end{eqnarray}} 
 
\begin{document} 
 
%\preprint{HUTP-05/A0048} 
 
\title{Speed of Sound in String Gas Cosmology} 
 
\author{Nima Lashkari and Robert H. Brandenberger} 
\affiliation{Department of Physics, McGill University, 3600 University 
Street, Montr\'eal QC, H3A 2T8, Canada} 
 
\date{\today} 
 
\pacs{98.80.Cq} 
 
\begin{abstract} 

We consider an ensemble of closed strings in a compact space with stable one cycles
and compute the speed of sound  resulting from string thermodynamics. Possible 
applications
to the issue of Jeans instability in string gas cosmology are mentioned.

\end{abstract} 
 
\maketitle 
 
\section{Introduction} 
\label{sec:1} 
String gas cosmology \cite{BV} (see \cite{SGrevs} for some 
recent reviews)  is a scenario of the very early 
universe based on taking into account new degrees of freedom and 
new symmetries which characterize string theory but are absent in 
point particle field theories. In string gas cosmology (SGC), matter is 
treated as a gas of closed strings. According to SGC, 
the radiation phase of standard cosmology was preceded not by a 
period of inflation, but by a quasi-static Hagedorn phase during 
which the temperature of the string gas hovers close to the Hagedorn 
temperature \cite{Hagedorn}, the maximal temperature of a gas of closed 
strings in thermal equilibrium. Thus, it is hoped that the scenario 
will not have a singularity in its past. SGC provides 
a possible dynamical explanation \cite{BV} of why there are only three large 
spatial dimensions (see, however, \cite{Columbia,Danos} for some 
concerns), provides a simple and physical mechanism of stabilizing 
all of the size \cite{Patil2} (see also \cite{Watson,Patil1}) and  
shape \cite{Edna} moduli of the extra dimensions, leaving only the 
dilaton multiplet un-fixed. The dilaton, in turn, can be fixed \cite{Danos2}
by making use of non-perturbative effects like gaugino condensation.   
 
It has recently been suggested \cite{NBV,Ali,BNPV1}
that thermal fluctuations of a gas of closed strings on a toroidal space will, 
in the context of the background 
cosmology of the string gas scenario, generate an almost scale-invariant 
spectrum of adiabatic, coherent cosmological fluctuations, fluctuations 
which have all of the correct properties to explain the recent high precision 
observations of cosmic microwave background anisotropies . 
Thermal fluctuations are given by the specific heat capacity. 
For a string gas living in a toroidal space the heat capacity depends on the area of torus. 
The deviation from extensivity is, in fact, a result of the enormous number of winding 
modes that become excited close to Hagedorn temperature. This 
result suggests that the heat capacity $c_V(R)$ scales holographically as a function of
the radius $R$ of a volume embedded inside the total box, i.e. $c_V(R) \sim R^2$.
This leads to the scale-invariance of the spectrum of cosmological
fluctuations \cite{NBV,Ali,BNPV1}, and implements a concrete realization of the
argument that holography almost always leads to scale invariance of 
perturbations \cite{Banks}. SGC, in fact, produces a scale invariant
spectrum of scalar metric perturbations with a slight red tilt, again like 
what is obtained in most inflationary models. The spectrum of gravitational waves, on 
the other hand, is characterized \cite{BNPV2} by a spectrum with a slight blue tilt, 
unlike the slight red tilt which is predicted in inflationary models. This
yields a way to observationally distinguish between the predictions of SGC
and inflation \cite{Stewart}.

In order to develop string gas cosmology into a viable alternative to 
inflationary cosmology, further issues need to be addressed. How are 
the horizon, flatness, size and entropy problems of Standard Big Bang 
cosmology addressed in string gas cosmology? What explains
the overall isotropy of the cosmic microwave background (horizon
problem)? why is the universe so large (size problem) and contains so
much entropy (entropy problem) compared to what would be expected
if the universe emerges from the Big Bang with a scale commensurate
with its initial temperature? Recall that it was these
questions which motivated the development of inflationary cosmology \cite{Guth}.

If SGC is embedded in a bouncing universe cosmology, as can be realized 
\cite{Biswas} in the ghost-free higher derivative gravity theory discussed in \cite{Siegel},
then the horizon, size and entropy problems do not arise
\footnote{See \cite{Natalia} for an approach to resolving the size problem
without assuming a bouncing cosmology.}. The flatness
problem, however, persists. In this Note we focus on 
the flatness problem. The flatness problem has two aspects: firstly, what 
explains the overall nearly spatially flat geometry of the current universe, 
and, secondly, why there are no large amplitude
small-scale fluctuations which will 
collapse into black holes and prevent the SGC scenario from working. It is this 
second aspect of the flatness problem which will be addressed in this Note (the
first one is again not present if the Hagedorn phase of SGC occurs around the
time of a cosmological bounce and we assume that at some point in the
contracting phase the universe was as large as it is today and the spatial
curvature is comparable to the current spatial curvature).

In cosmology, the Jeans length determines whether small scale structures
can collapse. The Jeans length, in turn, is determined by the speed of sound.
If the speed of sound is of the order unity (in units in which the speed of
light is unity), then the Jeans length, the length below which perturbations
are supported against collapse by pressure, is given by the Hubble length
$H^{-1}$, where $H$ is the expansion rate of space (see e.g. \cite{MFB}
for a review of the theory of cosmological fluctuations and \cite{RHBrev}
for an overview). In a radiation-dominated universe the speed of sound
is $c_s^2 = 1/3$ and hence small-scale instabilities do not occur. In
contrast, in a matter-dominated universe $c_s = 0$ and Jeans instability
on small scales occurs. It is this gravitational instability which leads to
the formation of nonlinear structures in the present universe.

In this Note we compute the speed of sound in the Hagedorn phase of
SGC. Matter in the Hagedorn phase is dominated by relativistic
strings containing both momentum and winding modes. On this
basis, we might expect that locally flat space is protected against
the Jeans instability. On the other hand, the net pressure is small
since the positive pressure of the momentum modes is cancelled by
the negative pressure of the winding modes. This leads to the
expectation that the speed of sound might be small.

In this Note we show that the speed of sound calculated using 
the microcanonical ensemble is very small and positive. As soon as 
the radius of torus becomes two orders of magnitude or so larger than 
the string length, the speed of sound goes to zero exponentially with 
the radius. On this basis, it appears that SGC may suffer from a
small-scale Jeans instability problem, since, in order to make
contact with the current universe, the radius $R$ needs to be at least
1 mm (if the string scale is comparable to the Grand
Unification scale). However, since the microcanonical ensemble
has limited applicability in describing the physics of subspaces
of the entire space, we need other techniques in order to
unambiguously resolve the Jeans instability puzzle of SGC.

There has been some previous work on inhomogeneities in the early
phases of SGC. Specifically, in the context of taking
the background of string gas cosmology to be described by dilaton
gravity, as in \cite{TV,BEK,Borunda}, it has been shown \cite{Watson0}
that there is no growth of cosmological inhomogeneities. 
Dilaton gravity cannot, as is now
realized \cite{Betal,KKLM}, provide a consistent background for the
Hagedorn phase of string gas cosmology because of the rapid time
variation of the dilaton. Since the dilaton velocity was playing
an important role in the considerations of \cite{Watson0}, we have
to revisit the issue of stability towards growth of fluctuations.

In the following section, we will review the microcanonical approach to 
string thermodynamics. We discuss the physics of an ideal gas of 
strings, the equilibrium conditions and the energy distribution in the gas. 
In Section 3 we will focus on the microcanonical ensemble to find the 
speed of sound for string gas fluctuations. In the absence of a well-defined 
canonical ensemble, the interpretation of the speed of sound becomes subtle. 
The final section contains a discussion of the interpretation of results and 
conclusions.

\section{Thermodynamics of String Gas}

There are two independent approaches to thermodynamics of string gas. 
The microcanonical ensemble approach starts with single string density of states 
calculated from the tower of states of a free string. Using this, one can find the 
number distribution of strings as a function of their energy. In this approach one 
finds the entropy of the microcanonical ensemble assuming equi-partition \cite{DJT}. 
The second approach starts with Boltzmann equations and derives the equilibrium 
distribution of strings using the canonical ensemble \cite{LT}. The main advantage 
of the second approach is that it can study the time dependent properties of string gas.
However, the canonical ensemble breaks down as the temperature reaches 
the Hagedorn temperature, rendering this approach useless for our purpose. 
Consequently, here we will make use of the first approach.

In spite of the fact that the canonical ensemble is not well-defined, one can still 
make sense of the partition function $Z(\beta)$ as a mathematical function, namely as
the Laplace transform of the density of states $\Omega(E)$:
\bea\label{partitition}
Z(\beta) \, = \, \int^\infty_0dE e^{-\beta E}\Omega(E)
\eea
where $\beta$ is the inverse temperature.
In the microcanonical ensemble approach, one employs the knowledge of 
the string spectrum to  construct the partitition function. Then, the density of states 
can be found by taking the inverse Laplace transform. The partition
function has singularities as values corresponding to distinct
temperatures larger than the Hagedorn temperature. The position of the 
singularities of the partitition function depends on the radius of compactification.
As  the torus grows in size, the singularities move closer to the 
Hagedorn temperature which we denote by $\beta = \beta_0$ \cite{DJT}. 
The  contribution of each singularity in 
the partition function to the density of states becomes smaller the further it is 
from the Hagedorn temperature. 

As in previous work on SGC, we will work with the density of states $\Omega$
for a gas of strings in nine spatial dimensions, of which three are large 
(but compact). The reason for choosing compact rather than non-compact
topology for our three large spatial dimensions is that string
thermodynamics is better defined than in the non-compact case.
If space is not compact, one does not obtain the holographic scaling
of the specific heat capacity which is required to obtain a scale-invariant
spectrum of cosmological perturbations. Also,  it has been shown that 
in the context of a string gas in thermal equilibrium in three or more 
non-compact dimensions,  at high energy densities a single energetic string 
captures most of the energy in the gas. More problematically, the specific 
heat turns out to be negative in this topology \cite{DJT}. 
%Nine string length compact dimensions can not have important cosmological 
%implications either. Satisfying equilibrium requires a not too small string coupling, 
%however a too large coupling simply forms black holes. Therefore, there is a 
%compromise and one has to be careful with the expectation value of Dilaton.\cite{Danos}.

String thermodynamics tells us how the energy is being distributed among strings. 
If one defines $D(\epsilon,E)$ to be the average number of strings that carry 
energy $\epsilon$ while the whole gas has overall energy $E$, the energy 
distribution of the string gas in three compact but large dimensions 
follows Fig. 1 \cite{DJT}. The absence of a peak which would correspond to the most 
probable value is the reason that assigning a mean value to a sub-box as we do in 
the canonical ensemble is not valid anymore. This means that although one can 
still calculate the partition function for a string gas, 
the canonical ensemble loses its ability to determine the thermodynamical quantities 
of a sub-box of the total box. Therefore, we have to restrict ourselves to 
the microcanonical ensemble. The physical way of understanding this is to notice that 
the thermodynamics of a string gas starts deviating from that of a gas of point particles 
as soon as there is enough energy to excite winding modes 
\footnote{The thermodynamics of a gas of open strings is extensive and resembles 
that of a gas of point particles due to the absence of winding modes.}. 
Winding modes are not localized degrees of freedom and result in non-extensivity 
of the entropy function (\ref{entropy}). Non-extensivity implies that erecting a wall 
anywhere in the box will cause bulk effects, unlike in the case of a gas of point 
particles where it is just a surface effect \cite{DJT}. 

In the next section we will calculate all the thermodynamical parameters that can 
be found using the microcanonical ensemble. Namely we will find the pressure, 
the temperature and the speed of sound for a gas of strings in a 
nine dimensional torus. 

\begin{figure}
\includegraphics[height=5cm]{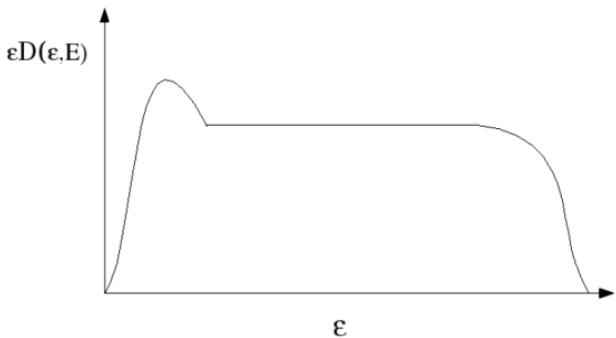} 
\caption{Energy distribution for a string gas in three large compact dimensions.} 
\label{fig:1} 
\end{figure}

\section{Speed of Sound of a String Gas}

We will start with three large dimensions as a three-dimensional
torus with equal radii $R$ \footnote{What is essential for the
analysis is the presence of one cycles on the manifold which
guarantees the existence of stable winding modes.}. As
has been derived in \cite{DJT}, the density of states for
fixed total energy $E$ and fixed $R$ and for energy densities above 
the Hagedorn energy density can be written as
\begin{equation} \label{density} 
\Omega(E,R) \, \simeq \, \beta_0e^{\beta_0E+aR^3}
\big [1+\delta\Omega_1\big ] \, , 
\end{equation} 
where
\begin{equation} \label{density2}
\delta\Omega_1 \, = -\, 
\frac{(\beta_0E)^5}{5!}\,e^{-(\beta_0-\beta_1)(E-\rho_0R^3)} 
\end{equation}
In string length units, $\beta_0$ the inverse Hagedorn temperature is simply 
$4\pi$ and $\beta_1$ is a constant smaller than $\beta_0$ which corresponds 
to the location of next singularity in the partition function \cite{DJT} and 
$\rho_0\sim \, l_s^{-4}$. In the above, $a$ is constant 
(of the order $l_s^{-3}$) due to the fact that the 
momentum density is almost constant above the Hagedorn energy density. 

The reason that we are interested in energy densities beyond Hagedorn 
energy density is that for lower values of energy density a gas of point 
particles becomes entropically favored and string gas acts effectively as a 
gas of low energy supergravity fields due to the lack of energy for exciting 
high oscillatory and winding modes. 
The ratio of the energy densities of a gas of closed strings to that of a gas of
point particles is given by
\be\label{rho}
\frac{S_{strings}}{S_{point-particles}}\sim \frac{\beta_0\rho}{\rho^{d/(d+1)}}\sim \Big(\frac{\rho}{\rho_0}\Big)^{1/4}
\ee

It can be shown that for large dimensions $R\gg l_s$ \cite{DJT}
\begin{equation} \label{approx} 
    \beta_0-\beta_1 \, \sim \, \frac{l_s^3}{R^2} \, , 
\end{equation}

Since the entropy is the logarithm of the density of states, we obtain 
\be \label{entropy} 
    S(E,R) \,  =  \, 
\ln{\beta_0} + \beta_0E + a\,R^3 +\ln(1+\delta\Omega_1) \, .
\ee
% Discussion of non-exitensivity of entropy
%
Knowing the entropy as a function of energy and radius we
immediately obtain the followiing formula for the temperature
\bea \label{temp} 
    T(E,R) \, &=& \, \bigg[\bigg (\frac{\partial S}{\partial E}\bigg )_V\bigg]^{-1} \\ 
    &=& \,  \bigg [\beta_0+\frac{1}{1+\delta\Omega_1}\frac{\partial\delta\Omega_1}{\partial E}\bigg ]_V^{-1} \, . \nonumber
\eea
Taking the derivative of (\ref{density2}) and  keeping in mind that $\rho \gg\rho_0$ 
we get
\be
  \frac{\partial\delta\Omega_1}{\partial E} \, = \, -(\beta_0-\beta_1)\delta\Omega_1 \, .
\ee
Inserting this result into (\ref{temp}) and making use of
(\ref{approx}) we obtain
\begin{equation} \label{deltaom} 
    \delta\Omega_1 \, = \, (\beta_0-\beta)\frac{R^2}{l_s^3} \, . 
\end{equation}
For simplicity we define the large positive parameter
\be\label{tau}
\tau=l_s\,(\beta-\beta_0)^{-1}
\ee
which is a measure of how close the system is to Hagedorn phase transition. 

Inserting (\ref{deltaom}) into  (\ref{density2}) and taking the logarithm of
both sides results in
\begin{equation} \label{energy} 
    E \, \sim \, 
\frac{R^2}{l_s^3}\bigg [\ln{\tau}+13\ln{\frac{R}{l_s}}+5\ln(l_s^{-4}\rho)+\ln{\frac{(\beta_0/l_s)^5}{5!}}\bigg ] \, . 
\end{equation}

The consistency of having $\delta \Omega_1 \ll 1$ implies, using (\ref{deltaom}),
that in large radius limit $R\gg l_s$ we have $\tau\gg (R/l_s)^2$. 
Thus, for the approximations made in the current analysis of string gas 
thermodynamics in three compact large dimensions to be consistent, the temperature 
has to be extremely close to the Hagedorn temperature. 
% lower temperatures, the string gas is almost just a gas of supergravity fields. 
In fact, for $R\gg 100 l_s$, the first term in (\ref{energy}) dominates, resulting in 
a more severe constraint on the temperature, namely $\tau\sim \exp(\rho Rl_s^3)$.

From our expression for the entropy, the pressure $p$ for adiabatic perturbations 
can be found by taking the derivative with respect to volume at constant energy
and making use of (\ref{deltaom}) to substitute for $\delta \Omega_1$. The
result is 

\begin{eqnarray}\label{pressure} 
    p \, &=& \, T\bigg (\frac{\partial S}{\partial V}\bigg)_E \\
&=& \, \frac{a}{\beta_0}-\bigg[a/\beta_0+(2\rho+\rho_0)/3\bigg]\tau^{-1}\,+\,O(\tau^{-2})
\nonumber 
\end{eqnarray}
Note that the pressure close to Hagedorn phase transition has a maximum value of
$aT_0$ at the Hagedorn temperature and declines very slowly with
decreasing temperature. For energy densities much larger than the Hagedorn
temperature, the magnitude of the pressure is negligible in comparison to
the magnitude of the energy density. This result is expected since the
positive pressure of the momentum string modes is cancelled by the negative
pressure of the winding modes. It is the string oscillatory modes which are
responsible for the net positive. 

Taking the partial derivative of (\ref{entropy}) with respect to the energy density 
at constant entropy one finds
\be\label{dpdrho}
\bigg(\frac{\partial R}{\partial\rho}\bigg)_S \, = \, -\frac{\beta_0R}{3(\beta_0\rho+a)}\,+\,O(\tau^{-1})
\ee
Taking the derivative of (\ref{energy}) and inserting (\ref{dpdrho}) therefore 
yields the following expression for the change of $\tau$ as one 
increases the density at constant entropy:
\be\label{dtaudrho}
\bigg(\frac{\partial\tau}{\partial\rho}\bigg)_S \,= \, 
\bigg[Rl_s^3-\frac{5}{\rho}-\Big(l_s^3\rho-\frac{13}{R}\Big)\frac{\beta_0 R}{3(\beta_0\rho+a)}\bigg]\tau\,+\,O(1)
\ee
 
The sound speed of the string gas can now be found by taking the derivative of (\ref{pressure})
with respect to energy density at constant entropy and inserting (\ref{dtaudrho}). The
result is
\bea\label{sound} 
    c_s^2 &=& \bigg(\frac{\partial p}{\partial \rho}\bigg)_S \nonumber\\ &=&\Big[Rl_s^3-\frac{5}{\rho}-\frac{(l_s^3\rho-\frac{13}{R})\beta_0R}{3(\beta_0\rho+a)}\Big]\Big[\frac{a}{\beta_0}+\frac{2\rho+\rho_0}{3}\Big]\frac{1}{\tau}\nonumber\\
&&-\frac{2}{3}\frac{1}{\tau}+O(\tau^{-2})
\eea
The most important things to learn from the above result are firstly that for large enough 
values of $R$ the speed of propagation of fluctuations in the string gas background 
in a box of radius $R$ at temperature $T$ and energy density $\rho$ is positive, and
secondly that it is extremely small in magnitude. The suppression of the magnitude is
related to the fact that the pressure is suppressed relative to the energy density.
In summary, as our most important results we have shown that 
\be\label{speed}
0 \, < \, c_s^2 \, = \, \frac{4}{9} \rho R l_s^2 (\beta-\beta_0) \, \ll \, 1 \, .
\ee
In order for the approximations made use of in our derivation to be valid, 
the temperature has to be close to the Hagedorn phase transition temperature 
with an accuracy better than one part in $\exp(\rho Rl_s^3)$.  
Thus, the speed of sound has the same behavior as it would in a very thin
medium.

One can also calculate the speed of sound when the radius is kept constant instead of 
the entropy. The answer turns out to have the same $\rho R\tau^{-1}$ suppression.
\be\label{speed2}
0 \ ,< \, c_R^2 \, = \,\frac{2\rho Rl_s^2}{3}(\beta-\beta_0) \, \ll \, 1 \, ,
\ee
which confirms the analogy with a very thin medium.
 
\section{Discussion and Conclusions}

In this note we have computed the speed of sound in a gas of closed
strings on a three dimensional torus, assuming that the toroidal
radii are equal. Our main result is that the speed of sound is
positive but highly suppressed when $R \gg l_s$. 

The positivity of the speed of sound has implications for the stability of 
cosmological perturbations.
In general relativity, for adiabatic fluctuations,
the scalar metric perturbation variable $\Phi$ (the fluctuation of the 
metric component $g_{00}$ in longitudinal gauge - the coordinate system in which
the metric is diagonal - whose physical meaning is that of
the relativistic generalization of the Newtonian gravitational potential) satisfies the
following equation of motion (see e.g. \cite{MFB} for a review of the theory
of cosmological perturbations)
\be \label{Phieq}
\Phi^{\prime \prime} + 3 {\cal H} \Phi^{\prime} - c_s^2 \nabla^2 \Phi
+ \bigl[ 2 {\cal H}^{\prime} + (1 + 3 c_s^2) {\cal H}^2 \bigr] \Phi \, = \, 0.
\ee
In the above, we have used conformal time $\eta$ which is related to the physical time
variable $t$ via the cosmological scale factor $a(t)$. 
\be
dt \, = \, a(t) d\eta \, .
\ee
The derivative with respect to
conformal time is denoted by a prime, and ${\cal H}$ is the scale factor in conformal
time. Neglecting the expansion of the universe, the solutions of (\ref{Phieq}) are
oscillatory if the speed of sound is positive:
\begin{equation} 
    \Phi_k(\eta) \, = \, Ae^{ikc_s\eta} + Be^{-ikc_s\eta}  \,  . 
\end{equation}
In particular, there is no instability of the metric fluctuations. 

On small scales, however, we should be concerned with the instability of density fluctuations.
On sub-Hubble scales and in a matter-dominated universe, the density
fluctuation $\delta \rho$ obeys the following equation of motion
\be \label{Newtdens}
\ddot{ \delta \rho} - c_s^2 \nabla^2 \delta \rho - 4 \pi G \rho_0 \delta \rho \, = \, 0 \, ,
\ee
once again neglecting both the expansion of space and the possible presence of
entropy fluctuations. This equation shows that fluctuations on scales larger than
the Jeans length (given by the wave number for which the second and the third term
in the above are equal) grow. If the speed of sound is comparable to the speed of
light, the Jeans length is of the order of the Hubble length $H^{-1}$ and there is
no Jeans instability problem. However, if the speed of sound is negligible (as in our
case) then there is a potential Jeans instability problem \footnote{Note that the expression for the speed of sound (\ref{speed}) is independent of the string coupling $g_s$. The string thermodynamics is defined in the limit of small $g_s$ the same way as in \cite{AW}. The Jeans length can be found using $R\geq \sqrt{\frac{\pi c_s}{g_s^2\rho}}$.}.

Since in the Hagedorn phase of string gas cosmology $p/\rho$ is
vanishingly small for large values of $R$, the equation of state is like that of a 
matter-dominated universe. However, since the basic objects which make up
the gas are not point particles but extended relativistically moving strings, it is
unlikely that equation (\ref{Newtdens}) applies to describe matter fluctuations
in SGC. Putting these considerations together, we conclude that
our study has so far not resolved the concern
that SGC might suffer from a Jeans instability problem. 
  
The interpretation of speed of sound in SGC is somewhat subtle. 
The reason is that in the thermodynamics of a string gas the equivalence of 
the microcanonical and canonical ensembles is lost due to the exponentially 
growing density of states close to the Hagedorn temperature. Thus, results
concerning the change in thermodynamical quantities as the size of the entire
sample box is changed cannot immediately be applied to questions
related to sub-boxes, like the question we are addressing here.

In standard thermodynamics the analogy between the canonical ensemble 
and the microcanonical ensemble  in the limit of large box sizes comes about 
because we can approximate the partition function -the Laplace transform of 
the density of states - by a saddle-point approximation, finding the average 
value of $E$ for any sub-system of interest with small fluctuations about the mean. 
However, if the density of states grows exponentially with energy, like 
in the case of strings, the saddle point approximation breaks down.
If one tries to push the canonical ensemble further, one obtains divergent 
fluctuations about the mean value.

Although non-extensive thermodynamics might seem very counter-intuitive, 
even classically in the presence of gravity it is somewhat inevitable. 
According to the ergodicity theorem, a system that evolves for a long 
time will be able to reach any small neighborhoud of a point in 
phase space. Therefore, there are trajectories
which run into regions in phase space that correspond to black holes. 
Once these black holes are nucleated out of thermal fluctuations, 
they grow and render the canonical ensemble ill-defined.
 
\begin{acknowledgments} 
 
This work is supported in part by an NSERC Discovery Grant, by the 
Canada Research Chair program, and by a FQRNT Team Grant. We wish
to thank Andrew Frey for valuable conversations. 
 
\end{acknowledgments}

\end{document}